\documentclass[letterpaper, 10 pt, journal]{IEEEtranTIE}
\IEEEoverridecommandlockouts   

\usepackage[noadjust]{cite}
\usepackage{amsmath,amssymb,amsfonts}
\usepackage{graphicx}
\usepackage{textcomp}
\usepackage{xcolor}
\usepackage{hyperref}
\usepackage{booktabs}
\usepackage{multirow}
\usepackage{tabularx}
\usepackage{courier}
\usepackage{float}
\usepackage{makecell}
\setcellgapes{5pt}
\usepackage{subcaption}
\usepackage{amsmath}
\usepackage{tikz}
\usepackage{mathdots}
\usepackage{yhmath}
\usepackage{cancel}
\usepackage{color}
\usepackage{siunitx}
\usepackage{array}
\usepackage{multirow}
\usepackage{amssymb}
\usepackage{gensymb}
\usepackage{tabularx}
\usepackage{booktabs}
\usetikzlibrary{fadings}
\usepackage{amssymb, amsmath, amsfonts}
\usepackage{  mathrsfs, color, bm,bbm}
\usepackage{ graphicx, epstopdf}
\usepackage{algorithm,  multirow,titlecaps,algorithmicx}
\usepackage[T1]{fontenc}

\usepackage{tikzsymbols}
\DeclareMathAlphabet{\mathbbold}{U}{bbold}{m}{n}

\usepackage[english]{babel}
\usepackage[utf8]{inputenc}
\usepackage[noend]{algpseudocode} 
\usepackage{amsmath,amssymb,amsfonts}
\usepackage{graphicx}
\usepackage{xcolor}
\usepackage{booktabs}
\usepackage{multirow}
\usepackage{courier}
\usepackage{float}
\usepackage{makecell}
\setcellgapes{5pt}
\newcommand{\ra}[1]{\renewcommand{\arraystretch}{#1}}

\newcommand{\argmin}{{\mathrm{argmin}}}

\newcommand{\GP}[2]{{\mathcal{G}\!\mathcal{P}}(#1,#2)}

\newcommand{\tr}{{\mathsf{T}}}

\iftrue

\else

\fi

\newcommand{\vc}[1]{{ \mathrm{#1} }}

\newcommand{\Dcal}{{\mathcal{D}}}

\newcommand{\Xcal}{{\mathcal{X}}}

\newcommand{\Rbb}{{\mathbb{R}}}

\newcommand{\vsd}{v_{\text{d}}} 
\newcommand{\vsq}{v_{\text{q}}} 
\newcommand{\isd}{i_{\text{d}}} 
\newcommand{\isq}{i_{\text{q}}} 
\newcommand{\Ls}{L_{\text{s}}}  
\newcommand{\Rs}{R_{\text{s}}}  
\newcommand{\Kb}{K_{\text{b}}}  
\newcommand{\Vqs}{V_{\text{q}}} 
\newcommand{\Kt}{K_{\text{t}}}  

\newcommand{\Kp}{K_\text{p}}       
\newcommand{\Kv}{K_\text{v}}       
\newcommand{\Ki}{K_\text{i}} 
\newcommand{\Tn}{T_\text{n}}       
\newcommand{\fs}{f^{\text{(s)}}}                
\newcommand{\Fs}{F^{\text{(s)}}}                
\newcommand{\Ns}{N^{\text{(s)}}}                
\newcommand{\hs}{h^{\text{(s)}}}                
\newcommand{\hsu}{\bar{h}^{\text{(s)}}}         
\newcommand{\Ts}{T^{\text{(s)}}_{90}}           
\newcommand{\es}{e^\text{(s)}}                  
\newcommand{\esSS}{e^\text{(s)}_{\text{ss}}}    
\newcommand{\gammas}{\gamma^\text{(s)}}         
\newcommand{\alphas}{\alpha^\text{(s)}}         
\newcommand{\fp}{f^{\text{(p)}}}                
\newcommand{\Fp}{F^{\text{(p)}}}                
\newcommand{\Np}{N^{\text{(p)}}}                
\newcommand{\hp}{h^{\text{(p)}}}                
\newcommand{\hpu}{\bar{h}^{\text{(p)}}}         
\newcommand{\Tp}{T^{\text{(p)}}_{90}}           
\newcommand{\ep}{e^\text{(p)}}                  
\newcommand{\epSS}{e^\text{(p)}_{\text{ss}}}    
\newcommand{\epZERO}{e^\text{(p)}_0}            
\newcommand{\gammap}{\gamma^\text{(p)}}         
\newcommand{\alphap}{\alpha^\text{(p)}}         
\newcommand{\mBO}{m_\text{BO}}          

\begin{document}

\title{Performance-Driven Cascade Controller Tuning with Bayesian Optimization}


\author{\IEEEauthorblockN{
Mohammad Khosravi, Varsha Behrunani, Piotr Myszkorowski, Roy S. Smith, Alisa Rupenyan and John Lygeros}
\thanks{The authors are with the Automatic Control Laboratory, ETH Z\"urich. A. Rupenyan is also with inspire AG, 8092 Z\"urich, Switzerland. (e-mails: \{khosravm, bvarsha,	rsmith, ralisa, lygeros\}@control.ee.ethz.ch). P. Myszkorowski is with Sigmatek AG, piotr.myszkorowski@sigmatek.ch.
} 
}

\maketitle

\begin{abstract} 

We propose a performance-based autotuning method for cascade control systems, where the parameters of a linear axis drive motion controller from two control loops are tuned jointly. Using Bayesian optimization as all parameters are tuned simultaneously, the method is guaranteed to converge asymptotically to the global optimum of the cost. The data-efficiency and performance of the method are studied numerically for several training configurations and compared numerically to those achieved with classical tuning methods and to the exhaustive evaluation of the cost. On the real system, the tracking performance and robustness against disturbances are compared experimentally to nominal tuning. The numerical study and the experimental data both demonstrate that the proposed automated tuning method is efficient in terms of required tuning iterations, robust to disturbances, and results in improved tracking.

\end{abstract}


\section{Introduction}

Routine maintenance of mechatronic systems requires the periodic tuning of PID controllers of linear or rotational drives to counter the gradual decline of performance due to the increase of friction, loosening of mechanical components, or wear. The corresponding PI and PID gains are often set to conservative values to ensure operation for a broad range of loads or mechanical properties, focusing on disturbance rejection.
While gain autotuning is in principle possible, existing off-the-shelf autotuning routines often interfere with safety mechanisms in manufacturing systems and are often avoided.

In classical model-based controller design, a model of the plant is derived from first-principle methods or identified using experimental data, and then a controller is designed based on this model \cite{feedbackSys}. For example, symmetric optimum tuning \cite{kessler58}, and its extensions \cite{Preitl99} provide relations between the controller gains and the achievable performance in terms of settling time, phase margin, overshoot, and other relevant metrics. Model-based methods are data-efficient and tuning can be achieved with relatively few iterations. For cascade control systems however, multiple cycles of tuning might be required, and the achieved values of the controller parameters are mostly on the conservative side. Whenever first-principle methods cannot be used, optimization problems arise in obtaining the final controller, both in computation of the model and the controller, through the minimization of prediction error criterion, or criteria based on controller stability and transient smoothness. A controller that best fits the frequency response of the optimal controller is obtained by optimization or order reduction techniques \cite{Karimi2017}. Performance-based objective functions based on calculating the error signal between the system output and the input reference signal such as e. g. Integral Square Error (ISE), Integral Absolute Error (IAE) or Integral Time Absolute Error (ITAE), are often minimized to tune controller parameters using genetic algorithms \cite{daSilva2000} or Particle Swarm Optimization \cite{Qi2019}, and often require a large number of evaluations to ensure that a global optimum is found. Standard methods, such as the Ziegler-Nichols rule \cite{ziegler1942optimum} or relay tuning with additional heuristics \cite{hang2002relay, Kumar2015} are routinely used for the tuning in practice and will be used here as benchmark results in simulation.

Various data-driven methods for controller design have been proposed, e.g. virtual reference feedback tuning (VRFT) \cite{Campi2002, Campi2006}, data-driven inversion-based control \cite{Novara2018}, direct learning of linear parameter varying (LPV) controllers from data \cite{Formentin2016}, data-driven controller tuning in mixed-sensitivity loop-shaping framework. Data-driven tuning has been explored in an iterative learning approach in \cite{Prochazka2005}, and also combined with a learned process model in \cite{Radac2014}. While they find application, the methods often work under assumptions of linearity or time invariance for the plant or the closed loop system. Iterative learning approaches might require multiple iterations or interventions through external probing signals or specially designed references. Data-driven methods often do not allow for constraints in the input and output variables, and the choice of reference models is also dependent on prior knowledge of the system \cite{Bazanella2014}.

Bayesian optimization (BO) is used for hyperparameter tuning in high-dimensional machine learning models, but it has also gained attention in engineering and has been applied e.g., to control of quadrotors combined with learned dynamics \cite{Tomlin}, optimization of process set-up parameters \cite{Maier2019, Maier_2020gr,}, and  collision avoidance \cite{Andersson2016}. Bayesian data-driven approaches have been also proposed for the tuning of PID controllers and of nonlinear systems \cite{BayesOpt1, BayesOpt2,khosravi2019controller,khosravi2019machine}. These approaches are particularly interesting because of the inherent data-efficiency and flexibility of the Bayesian optimization (BO) method. BO in controller tuning, where stability is guaranteed through safe exploration, has been proposed in \cite{BayesOpt1}, and applied to robotic applications \cite{BayesOpt5}, and in process systems \cite{khosravi2019controller}. Constraint BO tuning is proposed in \cite{SafetyBO2020}.

We propose a performance-based BO approach for autotuning of the cascaded controller parameters of a linear axis drive where the controller parameters of interest are tuned simultaneously. The tuning problem is formulated as an optimization where the controller parameters are the variables that ensure a minimum in the cost defined through a weighted sum of performance metrics extracted from encoder signals. In this work, we restrict the range of the optimization variables to a limited set where the system is stable. We focus on overshoot and setpoint tracking errors as performance metrics, and demonstrate that the method ensures disturbance rejection as in the nominal tuning methods. Simultaneous tuning of all parameters eliminates the need for successive tuning trials for each control loop, while the method is guaranteed to converge to a global optimum due to the properties of Bayesian optimization. The rest of this paper is organized as follows: Section \ref{sec:problem_statement} introduces the general problem of performance-driven cascaded controller tuning in linear axis drives. Section \ref{sec:system_model} presents the plant model derived from first principles and system identification techniques. Section \ref{sec:num_results} introduces the performance-based approach with numerical results comparing the proposed tuning method with standard tuning methods, also with an evaluation of the performance metrics on a grid. Experimental validation is reported in Section \ref{sec:experimental}.



\section{Performance-Based Tuning}
\begin{figure}[t]
	\centering
	\includegraphics[width = 0.37\textwidth]{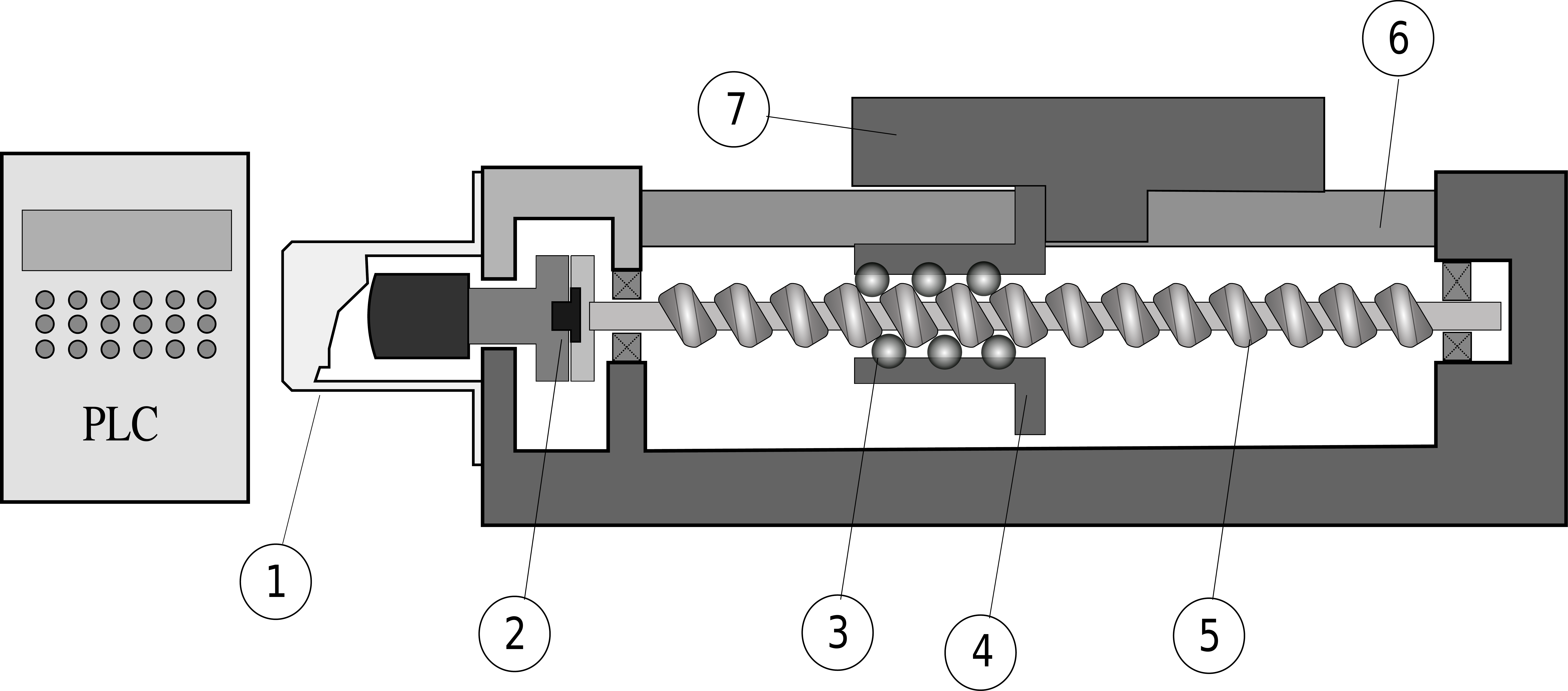}%
	\caption{The structure of ball-screw system:
		\textcircled{\small{1}} DC motor, \textcircled{\small{2}} coupling joint, \textcircled{\small{3}} ball-screw interface, 
		\textcircled{\small{4}} nut, 
		\textcircled{\small{5}} ball-screw shaft, \textcircled{\small{6}} guideway and 
		\textcircled{\small{7}} table (load), following \cite{altintas2011machine}.
	}
	\label{mfig:structure}
	\vspace{-4mm}
\end{figure}

\begin{figure}[h]
\vspace{-2mm}	\begin{subfigure}{.5\textwidth}
		\centering
		\includegraphics[width=.65\linewidth]{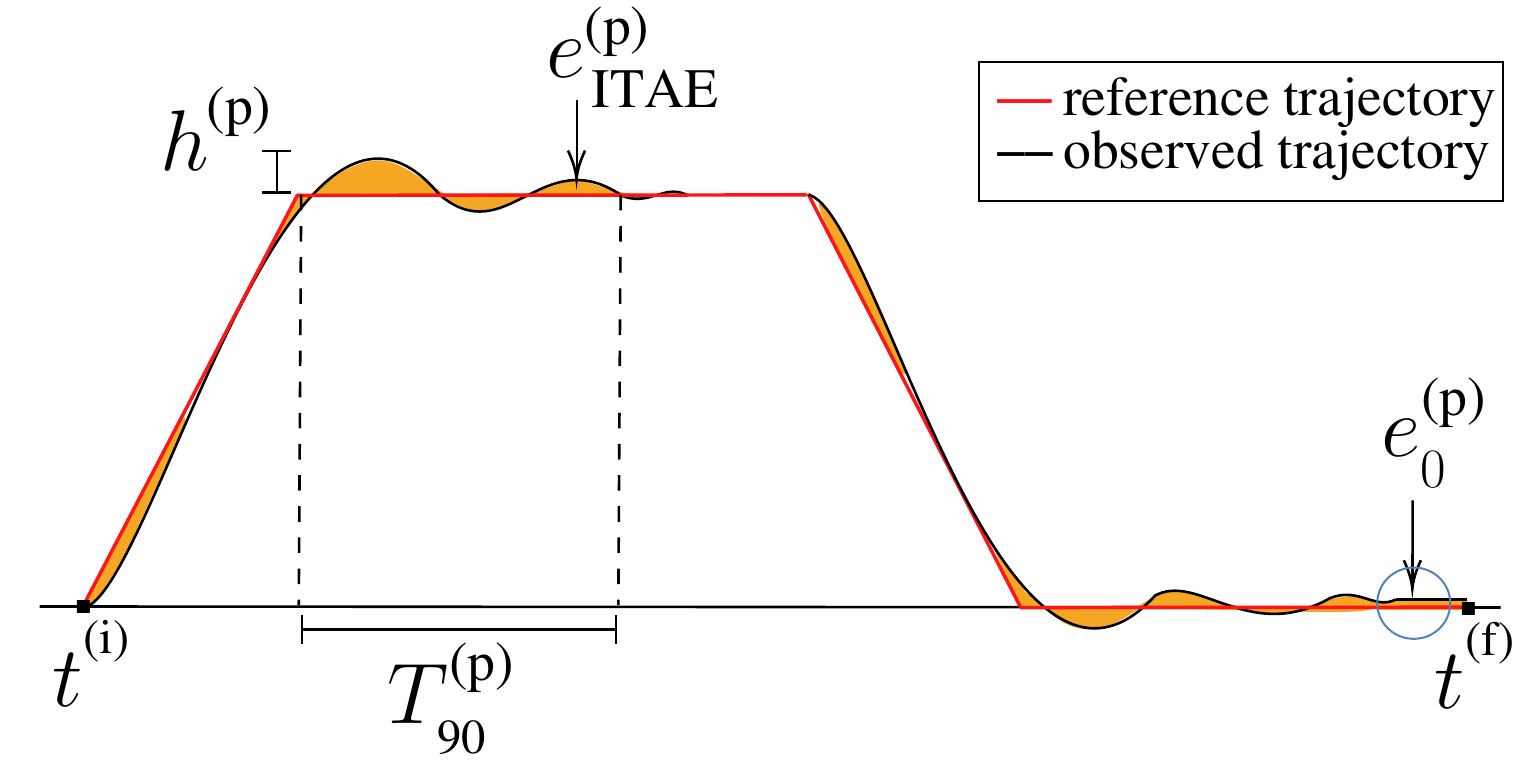}%
		\label{fig:perf_metric_posi}
	\end{subfigure}
    \vspace{0mm}\\
	\begin{subfigure}{.5\textwidth}
		\centering
		\includegraphics[width=.625\linewidth]{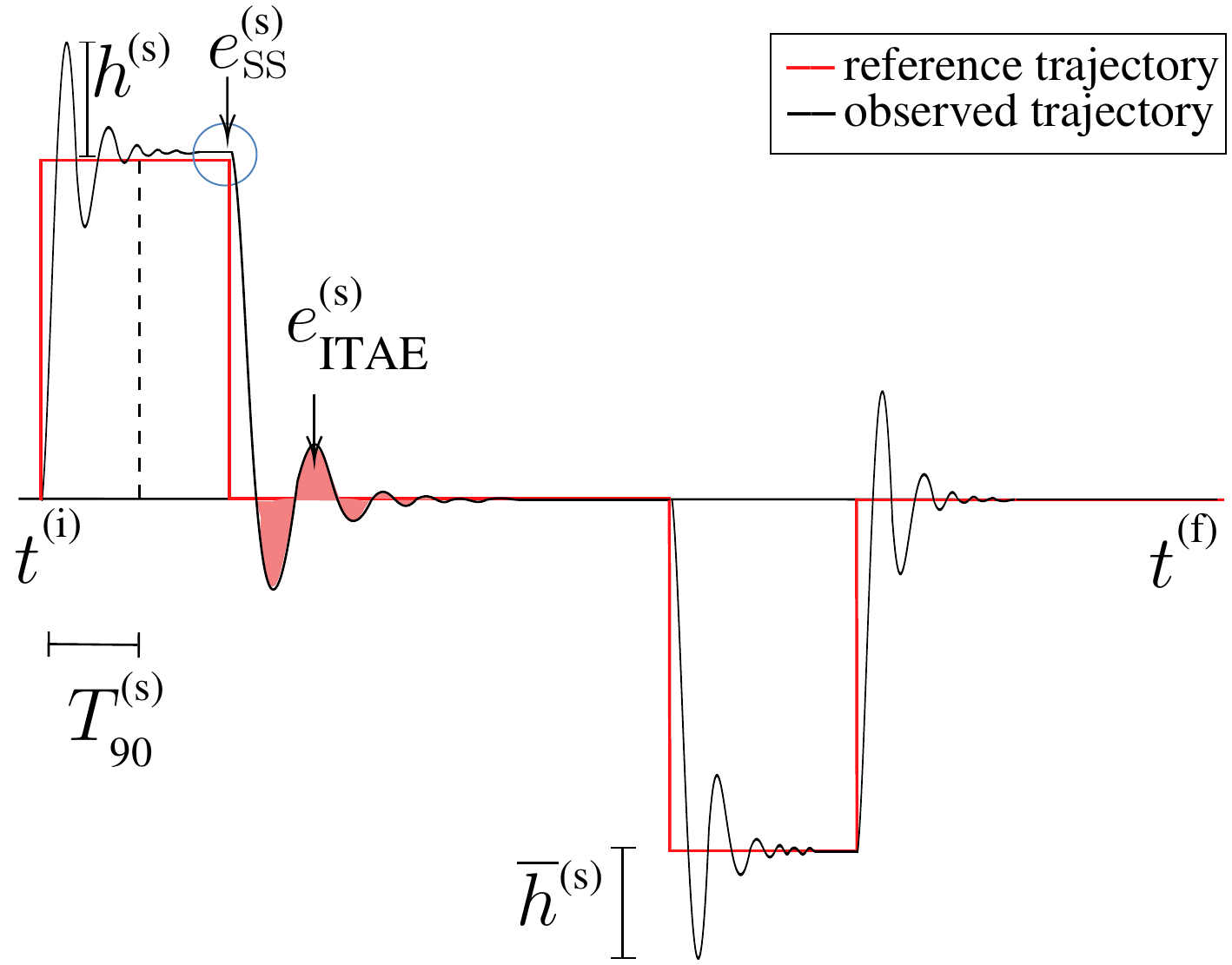}%
		\label{fig:perf_metric_speed}
	\end{subfigure}
	\caption{The figure demonstrates various performance metrics for position tracking (top) and speed tracking (bottom). The reference  and the observed trajectory are shown respectively 
	by the red line and
	black line.}
	\label{fig:perf_metrics}
\end{figure}

The system of interest is a ball-screw drive, a positioning mechanism used routinely in machining systems. Notable applications of such systems are in the semiconductor industry, in biomedical engineering, and in the photonics and solar technologies. The ball-screw drive components are shown in Figure \ref{mfig:structure}. The AC motor is connected via a coupling joint to a ball-screw shaft fixed to a supporting frame. 
\label{sec:problem_statement}
The shaft carries a nut which converts the rotational motion of the shaft linear motion of the nut through a screw-nut interface. 
The system is equipped with encoders for measuring the nut position, the rotational speed of the motor and the linear speed of the ball-screw shaft, and a sensor for measuring the motor current. The set voltage of the motor is controlled  through a motor drive and a programming logic controller (PLC). Typically, the motor rotates the shaft, the nut moves from initial to final position, remains there for some time and returns to the initial position. The position and speed of the nut are controlled such that they follow designed ideal reference trajectories (see Figure \ref{fig:perf_metrics}).

\begin{figure}[t]
\centering
\includegraphics[width = 0.37\textwidth]{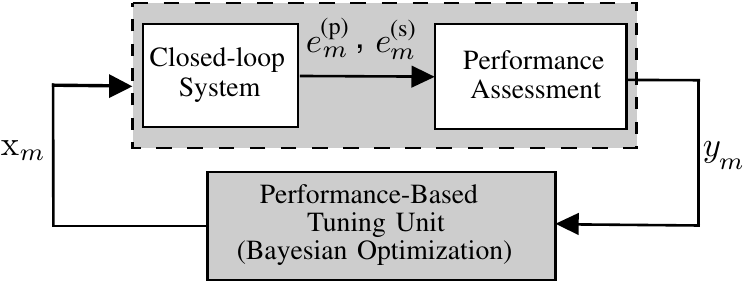}%
\caption{The performance-based controller tuning scheme.}
\label{fig:scheme}
\vspace{-4mm}
\end{figure}

 We introduce a performance metric indicating the quality of tracking, calculated from the position and speed error signals.  The error signals and the performance metric are functions of the vector of parameters describing the controller, denoted here by $\vc{x}$. Let the performance function be $f: \Xcal\to\Rbb$ where $\Xcal$ is the space of admissible controller parameters. 
With respect to controller parameters, $\vc{x}$, the performance function does not have a tractable closed-form expression, even when the dynamics of system are known. 
For a given $\vc{x} \in \Xcal$, the value of the performance metric $f(\vc{x})$  has a {\em black-box oracle} form, and can be obtained experimentally from the position and speed tracking error signals $\ep_m$ and $\es_m$, for controller parameters set to $\vc{x}$. 
We propose tuning the controller parameters based on a data-driven procedure of performing a sequence of experiments, collecting data, assessing the corresponding performance, and utilizing Bayesian optimization as a  black-box optimization technique for deriving optimal vector of controller gains (see Figure \ref{fig:scheme}). 
The estimated performance at each trial is provided to the performance-based tuning unit which is essentially a Bayesian optimization module. This unit models the performance metric using {\em Gaussian process regression} (GPR) \cite{GPR}, and finds the optimal parameters based on a suitable Bayesian optimization algorithm as {\em Gaussian process lower confidence bound} (GP-LCB) ~\cite{srinivas2012information}. 

The proposed tuning approach is  performance-based, model-free and data-driven. One of the main advantages of utilizing Bayesian optimization is its potential in explicitly modeling noise which is automatically considered in the uncertainty evaluation without skewing the result~\cite{srinivas2012information}. The simultaneous optimization of all parameters guarantees convergence to the global optimum performance, following from the properties of Bayesian optimization, at the cost of more iterations compared to classical approaches. 
\vspace{-0.5cm}

\section{System Structure and Model} \label{sec:system_model}

\subsection{Mathematical Model of System} \label{sec_plant_math}

The motor deriving the ball-screw system is a {\em permanent-magnet synchronous motor}, with a  permanent magnet on the rotor providing the exciting field. Let $\Rs$ and $\Ls$ respectively denote the resistance and the inductance of the stator.
From Kirchhoff's voltage law and the back electromotive force (EMF) which comes from Faraday's law of induction, in the dq-frame we have
\begin{equation*}
\label{eqn:v_dq_dynamics_1}
\!\!\!\!\!\!\!
\begin{split}
\vsd(t) &= 
\Ls \frac{\mathrm{d}}{\mathrm{d}t}\isd(t)
+ \Rs\isd(t) - \Ls\omega_{\text{m}}(t)\isq(t),
\\
\vsq(t) &= \Ls \frac{\mathrm{d}}{\mathrm{d}t}\isq(t)+ \Rs\isq(t) + \Ls\omega_{\text{m}}(t)\isd(t) + \Kb\omega_{\text{m}}(t),\\ 
\end{split}\!\!\!
\end{equation*}
where $[\vsd \ \vsq]^\tr$ and $[\isd\ \isq]^\tr$ are respectively the vectors of stator voltage and stator current in dq-coordinates, 
$\Kb$ is the back EMF constant and  $\omega_{\text{m}}$ is the angular velocity of the motor and shaft.
The d-component of the current, $\isd$, only influences the component of the magnetic field in the direction of the rotor axis and does not contribute to the torque generation. 
To minimize losses a PID-controller regulates $\vsd(t)$ to compensate the voltage induced by $\isq$ and ensures that $\isd = 0$ at all times.
Consequently, the resulting dynamics of motor is 
\begin{equation}\label{eqn:v_dq_dynamics_2}
\begin{split}
\vsd(t) &= - \Ls \omega_{\text{m}}(t)\isq(t) ,
\\
\vsq(t) &= \Ls \frac{\mathrm{d}}{\mathrm{d}t}\isq(t)+ \Rs\isq(t) + \Kb\omega_{\text{m}}(t).
\end{split}
\end{equation}
The motor develops an electromagnetic torque $\tau_{\text{m}} = \Kt \isq$ proportional to the stator q-component of current.
One can obtain the transfer function of motor as 
\begin{equation}\label{eqn:M(s)}
M(s) 
\!:=\! 
\frac{\Omega_{\text{m}}(s)}{\Vqs(s)}  
\!=\! 
K_{\text{t}}
\ \!
\big{(}K_{\text{t}}K_{\text{b}} \!+\!(\Ls s \!+\! \Rs)\Big(\frac{T_{\text{m}}(s)}{\Omega_{\text{m}}(s)}\Big)\big{)}^{-1}
\!\!,
\end{equation}
where $\Omega_{\text{m}}$, $\Vqs$ and $T_{\text{m}}$ are the Laplace transform of $\omega_{\text{m}}$, $\vsq$ and $\tau_{\text{m}}$, respectively.
The main limitation on the accuracy of the linear position is due to the {\em first axial mode} of the ball-screw system \cite{AxisModelling}, which is determined by the flexibility characteristics of the translating components.
This includes the inherent stiffness of the ball-screw and the bearings as well as the interactions between the carriage table, the nut and the ball-screw \cite{altintas2011machine,2DOFControl1,AxisModelling}.
The first axial dynamics of the ball-screw servo drive can be modelled using a simplified two degree of freedom  mass-spring-damper system \cite{altintas2011machine}.  
If we let $J_{\text{m}}$, $B_{\text{m}}$ and $\theta_{\text{m}}$ be inertia, damping coefficient and angular displacement of the motor, and $J_{\text{l}}$, $B_{l}$, $\theta_{\text{l}}$ and $\omega_{\text{l}}$ 
the corresponding quantities of the load, a torque balance leads to
\begin{equation}\label{eqn:omega_ml}
\begin{split}
&\!\! J_{\text{m}} \frac{\mathrm{d}\omega_{\text{m}}}{\text{d}t} + B_{\text{m}} \omega_{\text{m}} + B_{\text{ml}} (\omega_{\text{m}} - \omega_{\text{\text{l}}}) + K_{\text{s}} (\theta_{\text{m}}- \theta_{\text{l}}) = \tau_{\text{m}},\! \\&
\!\! J_{\text{l}}\ \! \frac{\text{d}\omega_{\text{l}}}{\text{d}t} + B_{\text{l}}\ \!\omega_{\text{l}} - B_{\text{ml}} (\omega_{\text{m}}- \omega_{\text{\text{l}}}) - K_{\text{s}} (\theta_{\text{m}} - \theta_{\text{l}}) = \tau_{\text{l}},\!
\end{split} 
\end{equation}
where $K_{\text{s}}$ is the total equivalent axial stiffness, $\tau_{\text{l}}$ is the torque disturbance of the load and $B_{\text{ml}}$ is the damping coefficient between the coupling and the guides. 
As $B_{\text{l}}$ has a negligible impact on resonance, one can set $B_{\text{l}}=0$ \cite{altintas2011machine}. 
Taking the Laplace transform leads to
\begin{equation}\label{eqn:Transfer_theta_tau}
\begin{bmatrix}
\Theta_{\text{m}}(s)\\\Theta_{\text{l}}(s)
\end{bmatrix}
=
\mathrm{H}(s)^{-1}
\begin{bmatrix}
T_{\text{m}}(s) \\ T_{\text{l}}(s)
\end{bmatrix},
\end{equation}
where capital letters denote the Laplace transform of the corresponding quantities and  $\mathrm{H}(s)$ is defined as
\begin{equation*}\label{eqn:H(s)}
\mathrm{H}(s)\! := \! \begin{bmatrix}
\!J_{\text{m}}s^2
\!+\!
(B_{\text{m}}\!+\!B_{\text{ml}})s
\!+\! 
K_{\text{s}}
& 
-B_{\text{ml}}s
-K_{\text{s}}
\\
- B_{\text{ml}}s - K_{\text{s}}
&\!\!\!\!\!\!\!
J_{\text{l}}s^2
\!+\!
(B_{\text{l}}\!+\!B_{\text{ml}})s
\!+\! 
K_{\text{s}}\!
\end{bmatrix}\!.
\end{equation*}
Due to the structure of the linear axial system, the torque disturbance of the load comparing to motor torque is negligible.
Therefore, to derive the transfer function from motor torque to the angular velocity, we set $T_{\text{l}}(s) =0$.
Let $D(s)$ denote the determinant of $\mathrm{H}(s)$. To obtain the transfer functions relating the motor to the angular velocities, we define transfer functions 
$F_1(s) := 
\Omega_{\text{m}}(s)/T_{\text{m}}(s) = (J_{\text{l}} s^2 + B_{\text{ml}} s + K_{\text{s}})/D(s)$, 
$F_2(s):= \Omega_{\mathrm{l}}(s)/T_{\text{m}}(s) = (B_{\text{ml}} s + K_{\text{s}})/D(s)$,
and subsequently,
\begin{align}
{F_3(s)} &:= \frac{\Omega_{\mathrm{l}}(s)}{\Omega_{\text{m}}(s)} = \frac{F_2(s)}{F_1(s)} = 
\frac{B_{\text{ml}} s + K_{\text{s}}}{J_{\text{l}} s^2 + B_{\text{ml}} s + K_{\text{s}}}. \label{eqn:T3}
\end{align}
Combining with \eqref{eqn:M(s)} leads to the
transfer function between the voltage applied to the armature and  rotational velocity of the load \cite{2DOFControl1}
\begin{equation} \label{eqn:G(s)_eq1}
\begin{split}
G(s) &:= \frac{\Omega_{\text{l}}(s)}{\Vqs(s)} = \frac{\Omega_{\text{m}}(s)}{\Vqs(s)} \frac{\Omega_{\text{l}}(s)}{\Omega_{\text{m}}(s)}= M(s)\ {F_3(s)}
\\&=
K_{\text{t}}
\ \!
\big{(}K_{\text{t}}K_{\text{b}} \!+\!(\Ls s \!+\! \Rs){F_1(s)}^{-1}\big{)}^{-1}{F_3(s)}.
\end{split}
\end{equation}
Moreover, if $K_{\text{s}} \gg 1 $, one can approximate $F_1(s)^{-1}$ by $((J_{\text{m}}+J_{\text{l}}) s +B_{\text{m}})$.
This approximation is valid for the low range of frequencies, typically $[0,10^5]$ Hz, which includes the main frequency range of the operation of the system. 
The final transfer function is 
\begin{equation}\label{eqn:11}
\begin{split}
\!\!\!\!
G(s) &=  
\Kt \bigg{(}\Kt \Kb + \big{(}\Ls s+ \Rs\big{)}\big{(}(J_{\text{m}}+J_\mathrm{l}) s +B_{\text{m}}\big{)}\bigg{)}^{-1}
\!\!\!\!\!
\\
& \qquad \qquad 
\bigg({\frac{B_{m\mathrm{l}} s + K_{\text{s}}}{J_\mathrm{l} s^2 + B_{m\mathrm{l}} s + K_{\text{s}}}}\bigg).
\end{split}
\end{equation} 

\subsection{The Control Scheme}
\label{sec:control}
\begin{figure}[b]
\vspace{-4mm}
	\centering
	\includegraphics[width=0.485\textwidth]{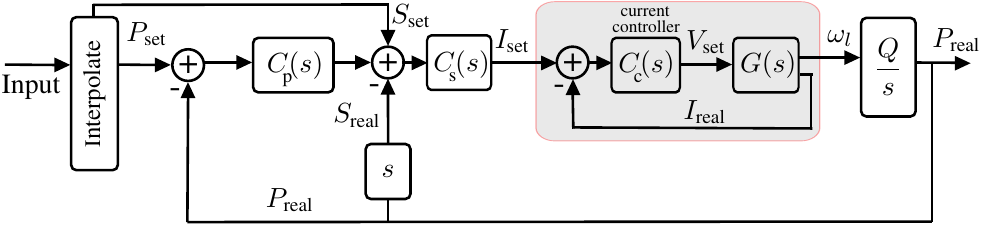}%
	\caption{Block diagram of the system}
	\label{fig:Block2}
\end{figure}

The system is controlled by a PLC that runs a custom-made software package named LASAL.
The controller consists of three cascaded loops as shown in Figure  \ref{fig:Block2}, where 
the output signals of each outer loop serves as the reference for the next inner loop. The first block in the axis controller is the interpolation block that receives the trajectory specifications from the user and determines the references for the position and the speed in the system. The interpolation block requires four inputs: position setpoint,  speed setpoint,  desired acceleration and desired deceleration. Once these inputs are provided, the interpolation block generates a reference speed and position trajectory using the equations of motion. The outer-most control loop is for linear position control. The middle control loop regulates the linear speed. 
The output of the interpolation block provides these loops with the designed nominal reference for the position and  the feed-forward reference for the speed. The motor encoder detects the position of the motor and  provides the feedback for both the position and speed control loops.
The controller in the position control loop is a P-controller as $C_\text{p}(s) = K_\text{p}$, whereas the controller in the speed control loop, is a PI-controller as 
$C_\text{s}(s) = K_\text{v} + K_\text{i}/s$.  
One can also introduce $C_\text{v}(s)$ in the form of 
$C_\text{s}(s)=\Kv (1+1/\Tn(s))$ where $\Tn$ is the integral time constant of the controller.  The speed control loop provides the reference for the current controller, which is the inner-most loop.
The feedback in this loop is the measured current of the armature.
This loop is regulated by a PID-controller block 
denoted by
$C_\text{c}(s) = K_\text{{cp}} + K_\text{{ci}}/s  + K_\text{{cd}} s.$ 
The output of the controller is the voltage setpoint for the motor given to the motor drive. 
Finally, the last block in Figure \ref{fig:Block2} is not part of the controller but is used to convert the rotational velocity of the ball-screw to linear speed. The factor $Q$ is the linear displacement resulting from one complete rotation of the motor.

The linear axis has three separate modes of operation according to which control loops and parameters are chosen. In position control mode (used in this work), all three feedback loops are active and the position is the critical attribute of the system. In this mode, the controller will try to
adhere as close as possible to the position reference even if that entails deviating from the speed trajectory. In the speed control mode (not considered here), the speed trajectory is prioritized,
and the position controller deactivated by setting the gain in
the position controller to zero, $\Kp = 0$. 
Finally, in the current control mode (also not considered here), only the inner-most loop is active and the other controller gains are set to zero.
\begin{table}[t]
	\caption{Parameters of the current controller and the plant.}
	\label{table:motor_Parameters}
	\centering
	\ra{1.3}
	\begin{tabular}[h]{@{}l c c c c c  @{}}\toprule
		\multicolumn{2}{c}{Controller $C(s)$}  &&   \multicolumn{2}{c}{Plant}\\
		\cmidrule{1-2} \cmidrule{3-5}
		Parameter  &  Value && Parameter  &  Value  \\ \midrule
		$K_\text{{cp}}$ & $60$ && $\Rs$  & $9.02$ $\Omega$	\\ 
		$K_\text{{ci}}$ & $1000$ 	&& $\Ls$  & $0.0187$   \\
		$K_\text{{cd}}$ & $18$ && $K_{\text{t}}$  & $0.515$ Vs $\text{rad}^{{-1}}$ \\ 
		&&& $K_{\text{b}}$  & $0.55$ Nm $\text{A}^{-1}$\\ 
		&&& $J_{\text{m}}$  & $0.27\times 10^{-4} \text{ kg m}^2$  \\
		&&& $B_{\text{m}}$  & $0.0074$  	\\ 
		&&& $J_\text{l}$ &  $6.53 \times 10^{-4}\text{kg m}{}^2$ \\ 
		&&& $B_\text{{ml}}$ &  $0.014$ 			\\
		&&& $K_\text{s} $ &  $3 \times 10^7$ 	\\ 
		&&& $Q$ &  $1.8\times 10^{-2}$ m 	\\ 
		&&& Max. Speed   & $8000$ RPM  	\\
		\bottomrule 
	\end{tabular}
\end{table}
\subsection{The Parameters of the Model}
The transfer function of the plant as well as the control loops depend on several parameters. 
As our work concentrates on tuning the parameters of $C_\text{p}(s)$ and $C_\text{s}(s)$, it is assumed that the parameters of $C_\text{c}(s)$ are given. 
Values for most of the parameters of the plant are provided or can be calculated from the available data sheets. The only exception is $K_{\text{s}}$. We estimate this parameter experimentally by fitting  the step response of the model using least squares. The resulting parameter values are summarised in Table \ref{table:motor_Parameters}.
\section{Performance-Based Controller Tuning: Numerical Investigation}\label{sec:num_results}
\subsection{Classical Tuning Methods} \label{sec:classical_tuning_methods}
The numerical study provided in this section elucidates various aspects and features of the proposed data-driven controller tuning method and compares its performance to widely used tuning techniques. The classical PID tuning approach is the Ziegler-Nichols method, a heuristic designed for disturbance rejection \cite{ziegler1942optimum}, and an automated method where the controller is replaced by a relay and the PID coefficients are estimated based the resulting oscillatory response of the system \cite{hang2002relay}.
Other tuning approaches are also used in practice, where a performance indicator of the system response is minimized, e.g., the {\em integral of time-weighted absolute error} (ITAE) \cite{aastrom1993automatic}.

\subsection{Performance-Based Tuning Method}
\label{sec:performance_based_tuning_methods}
The performance-based tuning proposed here utilizes Bayesian optimization to tune the controller parameters (see Figure \ref{fig:scheme}). 
The main ingredient in Bayesian optimization is the cost function, which is composed of a set of metrics capturing the performance requirements of the system. For a linear actuator, in addition to the standard controller tuning metrics such as overshoot and settling time, the position tracking accuracy and the suppression of mechanical vibrations known as oscillation effects are of highest importance. 

Let the vector of controller gains be set to $\vc{x} := [\Kp,\Kv,\Ki]^\tr$ and an experiment be performed in time interval  $I:=[t^{\text{(i)}},t^{\text{(f)}}]$ with these gains. Once the experiment is completed,  the corresponding error signals for position,  $\ep: I\to\Rbb$, and for speed, $\es: I\to\Rbb$, are measured, and used to calculate the tracking performance metric for $\vc{x}$. 


For position tracking, the corresponding performance metrics are the overshoot, $\hp$, the undershoot, $\hpu$, the steady-state error, $\epSS$, the settling time, $\Tp$, and the maximum absolute position error (infinity norm), $\| \ep\|_{\infty}$. To minimize oscillations due to the excitation of vibration modes, the integral of the time-weighted absolute value of the error of the position response also defined \cite{aastrom1993automatic} as
$\ep_{\text{ITAE}} := \int_{{t^{\text{(i)}}}}^{t^{\text{(f)}}} (t-t^{\text{(i)}})|\ep(t)|\ \mathrm{d}t$,
is  also included. 
The {\em position zero error}, $\epZERO$,  is defined as the error of the system once it returns to zero position (zero) once the motion is completed and it is of high interest, as it quantifies the oscillations in the system.
The same metrics are also used for the speed tracking error
and superscript `(s)' is used to denote the corresponding speed metrics.
\iffalse 
\begin{table}[t]
	\caption{Weights for the cost function of the position and the speed  controllers}
	\centering
	\ra{1.3}
	\begin{tabular}[h]{ @{}l c c c c c c c @{} }\toprule
		\multicolumn{2}{c}{position}  &&&   \multicolumn{2}{c}{speed}\\
		\cmidrule{1-4} \cmidrule{5-7}
		$\Fp_i$   & $\gammap$ &&&  $\Fs_i$ & $\gammas$ \\ \midrule 
		$\Tp$ & $10^4$              &&& $\Ts$ & $500$ &\\ 
		$\hp$ & $10$                &&& $\hs$ & $2$ & \\ 
		$\hpu$ & $0$                &&& $\hsu$ & $0$ & \\ 
		$\ep_{\text{ITAE}}$& $0$    &&& $\es_{\text{ITAE}}$ & $10^4$ & \\ 
		$\|\ep\|_{\infty}$ & $100$  &&& $\|\es\|_{\infty}$ & $500$ &\\
		$\epSS$ & $0$               &&& $\esSS$ & $0$ & \\ 
		$\epZERO$ & $0$             &&& && \\ 
		\midrule 
		$\alphap$ & $10$             &&& $\alphas$&$1$& \\ 
		\bottomrule
	\end{tabular}	
	\label{table:weights_grid}
\end{table}   
\else 
\begin{table}[b]
\centering
\vspace{-4mm}
	\ra{1.3}
	\begin{tabular}[h]{ @{}l c c c c c c c @{} }\toprule
		\multicolumn{2}{c}{position}  &&&   \multicolumn{2}{c}{speed}\\
		\cmidrule{1-4} \cmidrule{5-7}
		$\Fp_i$   & $\gammap$ &&&  $\Fs_i$ & $\gammas$ \\ \midrule 
		$\Tp$ & $10^5$              &&& $\Ts$ & $5\cdot 10^2$ &\\ 
		$\hp$ & $10^2$                &&& $\hs$ & $2$ & \\ 
		$\|\ep\|_{\infty}$ & $10^3$  &&& $\|\es\|_{\infty}$ & $5\cdot 10^2$ &\\
		& &&& $\es_{\text{ITAE}}$ & $10^4$ & \\ 
		\bottomrule
	\end{tabular}	
	\caption{Weights in the cost function 
	}\label{table:weights_grid}
\end{table}   
\fi

We take a weighted sum of the individual position metrics, denoted by $\Fp_i(\vc{x}), i=1,\ldots,\Np$, with corresponding weights $\gammap_i, i=1,\ldots,\Np$,  to generate a combined position metric $\fp(\vc{x})$. Similarly, we generate a combined speed metric, denoted by $\fs(\vc{x})$. In order to generate an overall performance metric for controller tuning, 
we then take sum of the two quantities as
$f(\vc{x}):=\fp(\vc{x})+\fs(\vc{x})$. 
The values of the different weights used are summarised in Table \ref{table:weights_grid}.
Each  of the weights depends on the scale and the importance of corresponding metric and feature and can be selected considering the particular application requirements, e.g. maximization of tracking accuracy in position, minimization of response time, or  minimization of oscillations in the system.

A fundamental constraint for tuning the controllers is the stability of the closed-loop system during the tuning procedure. This can be achieved by constraining the controller gains to known ranges guaranteeing this feature. 
To this end, the feasible set for $\vc{x}$, denoted by $\Xcal$, is defined as
\begin{equation}\label{eqn:X_gains_range}
\begin{split}
&[K_{\text{p},\min},\!K_{\text{p},\max}] 
\!\times\!
[K_{\text{v},\min},\!K_{\text{v},\max}] 
\!\times\!
[K_{\text{i},\min},\!K_{\text{i},\max}], \!
\end{split}
\end{equation}
where $K_{\text{p},\min}$,  
$K_{\text{v},\min}$ and  
$K_{\text{i},\min}$ are the smallest positive values acceptable by the controller,  
$K_{\text{p},\max} = 4200$,  
$K_{\text{v},\max} = 0.5$,  and
$K_{\text{i},\max} = 900$.
In practice, these ranges are either provided with the system, or estimated in a safe exploration procedure. Here, we have derived them by a grid computation of the system response for various controller parameters.

\begin{algorithm}[t]
	\caption{Performance-based Tuning Method}\label{alg:performace_based_tuning}
	\begin{algorithmic}[1]
		\State \textbf{Input:} Set $\Xcal$ (see \eqref{eqn:X_gains_range}), training data set, $\Dcal_0$, and trade-off weights $\gammap$, $\gammas$.
		\State Estimate the vector of hyperparameters $\hat{\theta}$.
		\State $\Dcal\leftarrow \Dcal_0$ and $m\leftarrow m_0$.
		\While{stopping condition is not met}
		\State Update the GP model: calculate $\mu_m$ and $\sigma_m$. 
		\State Derive $\vc{x}_{m+1}$ by solving \eqref{eqn:GP_LCB}.
		\State Update $\Kp$, $\Kv$ and $\Ki$, run experiment and measure signals $\ep_{m+1}$ and $\es_{m+1}$.
		\State Calculate the measured performance $y_{m+1}$.
		\State $\Dcal\leftarrow \Dcal \cup \{(\vc{x}_{m+1},y_{m+1})\}$ and $m\leftarrow m+1$.
		\EndWhile\label{DCP_loop}
		\State \textbf{end} 
		\State \textbf{Output:} $\vc{x}_{\mBO^*}$.
	\end{algorithmic}
\end{algorithm}
Note that $\fp$, $\fs$ and $f$ are functions of controller gains $\vc{x}$ in an {\em oracle form}.  The controller parameters with optimal performance, $\vc{x}^*$, are the ones inducing  minimum cost objective $f$. 
To find these optimal values, we utilize Bayesian optimization, summarized in Algorithm \ref{alg:performace_based_tuning}. 
Starting from feasible set $\Xcal$, we perform a sequence of experiments. 
In the $m^{\text{\tiny{th}}}$ experiment, a new vector of controller gains $\vc{x}_m$ is used, and based on the measured error signals, the corresponding performance 
$ f(\vc{x}_m)$ is evaluated as $y_m$.
The first $m_0$ experiments are for constructing the initial data set 
$\Dcal_0:=\{(\vc{x}_m,y_m)|m=1,2,\ldots,m_0\}$.  
Using GPR and the collected data, we obtain a surrogate function for the performance metric $f$ as a Gaussian process $\GP{\mu_m}{k_m}$, where $\mu_m:\Xcal\to\Rbb$ is the mean function predicting the value of $f$ at different locations, and $k_m:\Xcal\!\times\!\Xcal\to\Rbb$ is the kernel function evaluating the uncertainty in the predictions.
Here, we utilize {\em squared exponential} kernel with hyperparameters tuned  using $\Dcal_0$.
The controller gains are updated due to GP-LCB sampling algorithm \cite{srinivas2012information} as
\begin{equation}\label{eqn:GP_LCB}
\vc{x}_{m+1}=\argmin_{\vc{x} \in \Xcal}\ \mu_{m}(\vc{x})-\beta_{m}\sigma_{m}(\vc{x}) \, 
\end{equation}
where 
$\sigma_{m}(\vc{x}):=k_{m}(\vc{x},\vc{x})$ indicates the variance in prediction of performance function at $\vc{x}$ and 
$\beta_{m}$ is a constant specifying the considered confidence bound around the mean function $\mu_m$.
This procedure stops at iteration $\mBO$ either by reaching the maximum number of iterations, or by sampling around the same configuration of parameters with minimal cost more than three times.
The output of the algorithm is the vector of controller gains with the best observed performance, $\vc{x}_{\mBO^*}$.

One could also tune the gains of the two loops sequentially, alternating between the gains of the inner and the outer loop, utilizing BO for each step \cite{IFACBOCascade}. We do not pursue this direction further here, since the combined tuning approach introduced above is more likely to lead to a globally optimal solution.

\subsection{Numerical Experiments} \label{sec: BO_sim}
We now compare the system trajectories of the transfer function model from Section \ref{sec:system_model} corresponding to BO tuning, classical methods, and an exhaustive (grid) computation of the performance metric on a $280 \times 90 \times 100$ grid on $\Xcal$. 
To apply Algorithm \ref{alg:performace_based_tuning}  we initialize with dataset $\Dcal_0$ collected from random locations in $\Xcal$. In each iteration, we solve the GP-LCB optimization problem \eqref{eqn:GP_LCB} by calculating the objective function of \eqref{eqn:GP_LCB} for the points of the grid and finding the point with the optimal cost; note that this process does not require data to be collected at all the grid points, only that the Gaussian process is evaluated at these points.
Table \ref{table:SimCompare} shows that the controller gains, as well as the performance resulting from the proposed method are almost the same as the performance of optimal grid point, i.e., the performance-tuning method finds the controller gains in the nearly flat region of performance around the optimal gains.
\begin{table}[t]
\vspace{-1mm}		
\caption{Results of different tuning methods}
	\label{table:SimCompare}
	\centering
	\ra{1.3}
	\begin{tabular}[h]{@{}l c c c c c  @{}}\toprule
		Tuning method  &  $\Kp$ &  $\Kv$ &  $\Ki$ & $f$\\ \midrule
		Grid search (true optimal value)&225&0.39&90& 4557\\ 
		Ziegler Nichols &392&0.18&510 & 28434\\  
		ITAE criterion&255&0.11&420 & 31163 \\ 
		Relay Tuning&115&0.05&130& 23396 \\ 
		Performance-based Tuning &240&0.39&100& 4586\\ \bottomrule
	\end{tabular}
\vspace{-4mm}	
\end{table}

\begin{figure}[t]
	\centering
	\includegraphics[width=0.485\textwidth]{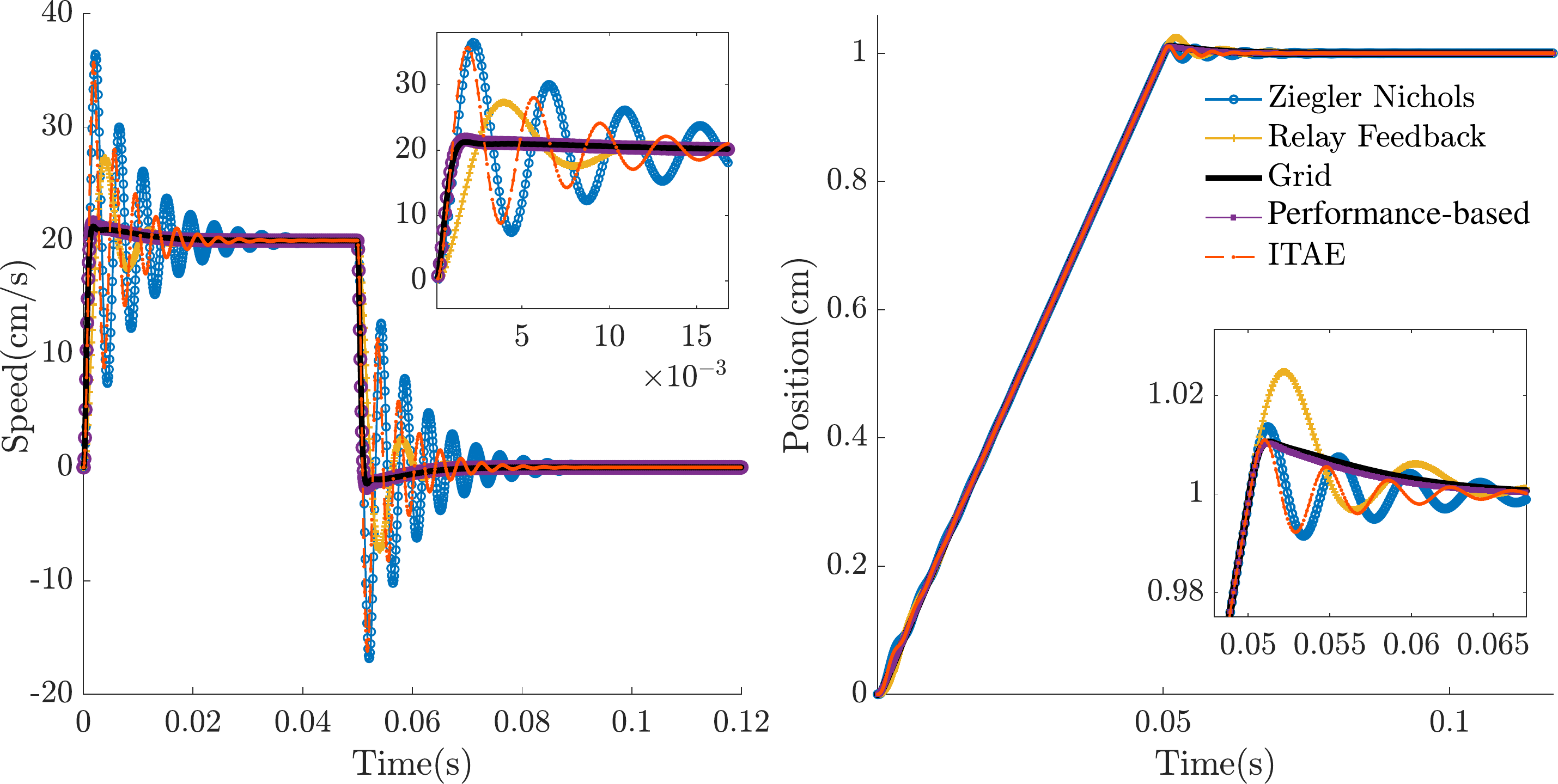}
	\caption{Response for different benchmark tuning methods}
	\label{fig:benchmark}
	\vspace{-6mm}
\end{figure}

The closed-loop speed and position responses corresponding to benchmark tuning methods to the performance-based tuning responses, namely Zeigler-Nichols method, relay tuning, and ITAE in Figure \ref{fig:benchmark} show significantly higher overshoot and oscillations compared to the performance-based tuning responses, which also matches the ideal response obtained by grid simulation. The position response tracks the input reference precisely, and the speed response has a slight overshoot in the acceptable range.
Table \ref{table:SimCompare} shows the resulting control gains and corresponding performance. The result of the performance-based tuning is closest to the exhaustive evaluation results obtained on the grid while the gains obtained via standard tuning approaches (Zeigler-Nichols, relay tuning, and ITAE) are more aggressive.
The number of training samples in $\Dcal_0$ has a direct influence on the number of iterations needed to reach the stopping criterion that defines the converged controller gains. A more informed prior model requires less iterations during the optimization phase.
We further performed a Monte Carlo simulation, where the algorithm was initialized with a different number of data points selected randomly from $\Xcal$, and for each value of $m_0$ Algorithm 1 was executed 1000 times for a maximum number of 60 iterations.
Based on these numerical experiments, 
an initial data set $\Dcal_0$ comprising 20-50 different random configurations of parameters requires a tuning phase of $20$ up to $50$ iterations in total for convergence, which is a reasonable trade-off between the number of initial samples and tuning phase iterations.
Since the performance metrics can be fully automated, and the initial exploration phase needs to be repeated only upon major changes in the system, the proposed tuning method can be efficiently implemented.
The numerical experiments are useful to clarify several open points in the performance-based autotuning, even though tuning on the real system does not require a model and is purely data-driven. First, the numerical experiments give an initial, safe indication if the method is working or not. Second, they indicate how many training points we should expect in realty to have a reasonable number of iterations for convergence. Similarly, they indicate the number of iterations need for convergence, and finding a suitable stopping criterion. Finally, they provide the range of stability in the controller gains. If the stability range is unknown, and if unstable parameters cannot be detected in advance, the application in reality might bring failures in the system. The method applied in practice is not influenced by the model or by the optimal parameters found in numerical experiments, as it provides the true optima according to the corresponding convergence criteria.
 
\section{Performance-Based Controller Tuning: Experimental Results} \label{sec:experimental}
The linear motion system (stage) used to validate the proposed tuning method consists of a linear axis, a permanent magnet AC motor with a servo drive, and a Programmable Logic Controller (PLC), as shown in Figure \ref{mfig:structure}. The system is equipped with a linear encoder with a precision of $1\mu\text{m/m}$ and sampling time of $1\text{ms}$ used for measuring the actual position and speed of the stage. The input voltage to the motor is provided based on the voltage reference signal from a Sigmatek S-DIAS PLC. 
Since the bandwidth of the current controller is above $1500\text{Hz}$ and the corresponding bandwidth of the position controller does not exceed $400\text{Hz}$, the current closed-loop can be well approximated with a constant gain. The proportional controller gains $\Kv$ and $\Kp$ of the cascade control loop are tuned, and the integral time constant $\Tn$ is tuned instead of the the integral gain $\Ki$.

For the joint tuning of the control parameters, the input setpoints (position, speed, acceleration, deceleration) were set identical for each run of the system. During the experiments, acceleration and deceleration values were kept high to simulate step input and step response in both position and speed control modes, and the speed was set at $20\frac{\text{cm}}{\text{s}}$. The range of the three control parameters was limited following the numerical experiments to ensure that the system is always  in a stable mode: $K_\text{p} \in (0,65000], K_\text{v} \in (0,7000], T_\text{n} \in (1000,40000]$.
\iffalse 
\begin{table}[t]
	\caption{Weights for the cost function of the position and the speed  controllers}
	\centering
	\ra{1.3}
	\begin{tabular}[h]{ @{}l c c c c c c c @{} }\toprule
		\multicolumn{2}{c}{position}  &&&   \multicolumn{2}{c}{speed}\\
		\cmidrule{1-4} \cmidrule{5-7}
		$\Fp_i$   & $\gammap$ &&&  $\Fs_i$ & $\gammas$ \\ \midrule 
		$\Tp$ & $20$                 &&& $\Ts$ & $20$ &\\ 
		$\hp$ & $50000$              &&& $\hs$ & $1000$ & \\ 
		$\hpu$ & $0$                &&& $\hsu$ & $2000$ & \\ 
		$\ep_{\text{ITAE}}$& $0$    &&& $\es_{\text{ITAE}}$ & $250000$ & \\ 
		$\|\ep\|_{\infty}$ & $50000$ &&& $\|\es\|_{\infty}$ & $0$ &\\
		$\epSS$ & $0$               &&& $\esSS$ & $500$ & \\ 
		$\epZERO$ & $10^5$          &&& && \\ 
		\midrule 
		$\alphap$ & $1$            &&& $\alphas$&$1$& \\ 
		\bottomrule
	\end{tabular}	
	\label{table:weights_grid_exp}
\end{table}
\else
\begin{table}[b]
\vspace{-4mm}	
	\centering
	\ra{1.3}
	\begin{tabular}[h]{ @{}l c c c c c c c @{} }\toprule
		\multicolumn{2}{c}{position}  &&&   \multicolumn{2}{c}{speed}\\
		\cmidrule{1-4} \cmidrule{5-7}
		$\Fp_i$   & $\gammap$ &&&  $\Fs_i$ & $\gammas$ \\ \midrule 
		$\Tp$ & $2\cdot 10^1$                 &&& $\Ts$ & $2\cdot 10^1$ &\\ 
		$\hp$ & $5\cdot 10^4$              &&& $\hs$ & $10^3$ & \\ 
		$\|\ep\|_{\infty}$ & $5\cdot 10^4$ &&& $\es_{\text{ITAE}}$ & $2.5\cdot 10^5$ &\\
		$\epZERO$ & $10^5$               &&& $\esSS$ & $5\cdot 10^2$ & \\ 
		&                &&& $\hsu$ & $2\cdot 10^3$ & \\ 
		\bottomrule
	\end{tabular}	
		\caption{Weights in the cost function
	}
	\label{table:weights_grid_exp}
\end{table}
\fi 

The position reference trajectory was a bi-directional step response, where the nut moves from  $0$ to $50$cm, remains there for $10$s, and returns to position $0$. In this case, input position and speed are set to $5$cm and $100\frac{\text{cm}}{s}$, respectively. 
The modified weights used in the experimental implementation of the performance-based tuning are provided in Table \ref{table:weights_grid_exp}. The termination criterion for tuning in practice was based on either achieving a predefined number of iterations, or reaching a minimum within a defined threshold and sampling at this candidate configuration for at least $3$ iterations.



\subsection{Experimental Results: Performance and Robustness}

\begin{figure}[h]
	\centering
	\includegraphics[width=0.425\textwidth]{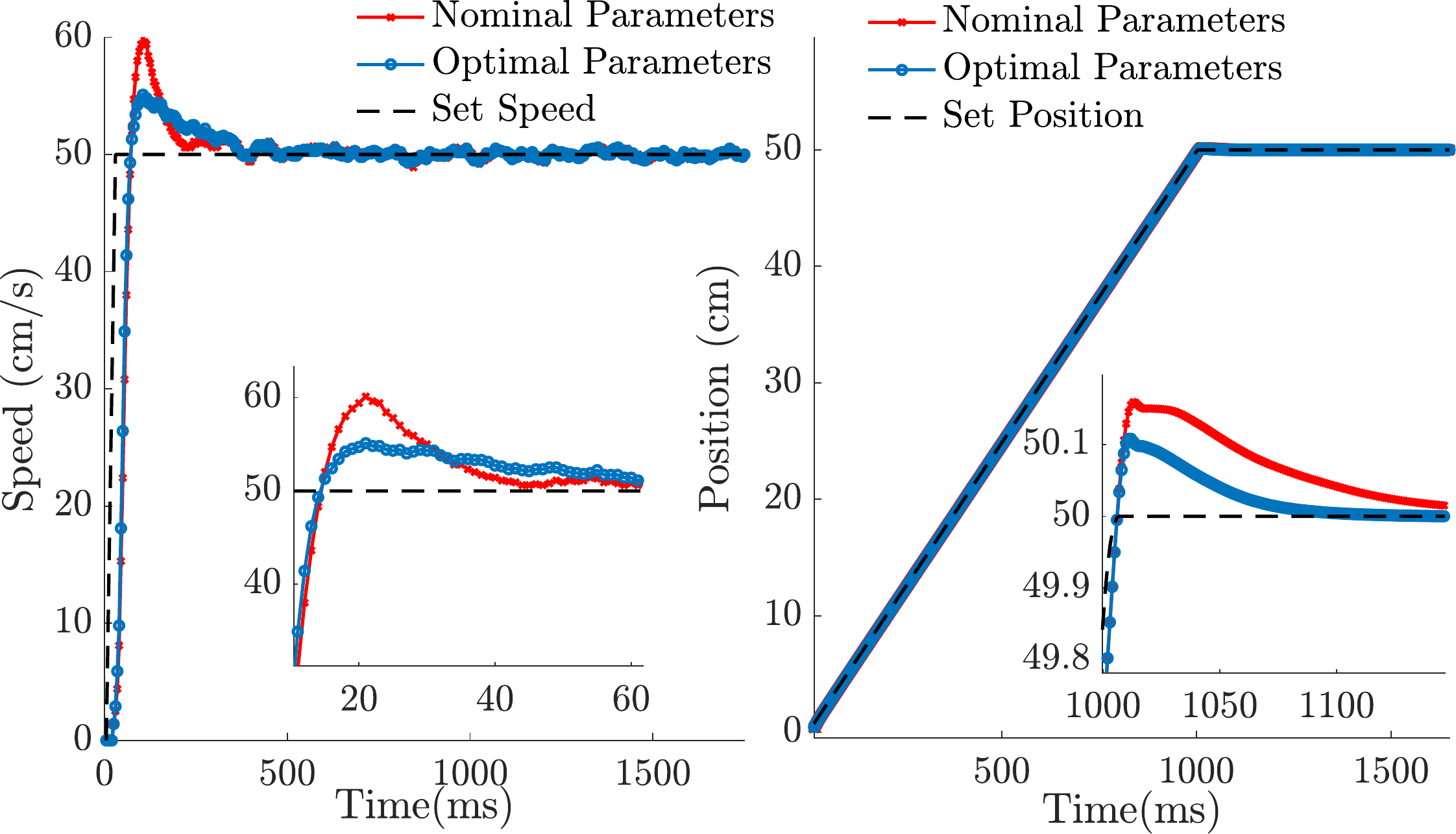}
	\caption{Position and speed responses comparing tracking traces of the system optimized with performance-based BO tuning and traces corresponding to the nominal parameters optimized by manual tuning. The insets show the initial overshoot obtained with each method.}
	\label{fig:ExperBO_grid}
	\vspace{-4mm}
\end{figure}

Initially, $49$ samples were obtained for different values of $\Kp$, $\Kv$ and $\Tn$. 
The required performance metrics were extracted from each of the data sets and the corresponding cost was calculated for each configuration of gains. 
Tuning the parameters using these training data for computing priors results in convergence after $18$ iterations. 
The optimal point is reached after repeated sampling near the optimum of the cost. The optimal values for the position and speed controller obtained are $\Kp = 44500$, $\Kv = 4000$ and $\Tn = 12500$. The system response in position control mode using the above controller configuration for set position of $50$cm and set speed of $50 \frac{\text{cm}}{\text{s}}$ is shown in Figure \ref{fig:ExperBO_grid}.
The negligible oscillation on the response is due to the amplified noisy resolver feedback, observed also in the nominal configuration. The speed response in the position control mode shown in Figure \ref{fig:ExperBO_grid} also has an overshoot of $14$\% which is higher than the speed control mode as expected but is within an acceptable range with a maximum speed of $57.5\frac{cm}{s}$. This is also in part due to the noise of the system. The position response steady state error is significantly smaller as compared to that in nominal mode, and the overshoot in position is less than $0.2$cm with very fast settling so there is no tracking delay or error.
 Performance-based tuning achieves better tracking and lower overshoot for both speed and position modes, and faster settling time for the more important position control mode. The results are summarized in Table \ref{table:expResults}, where it can be seen that the optimized parameters have reached higher values than the parameters from nominal tuning for the same load.

\begin{table}[b]
\vspace{-4mm}
	\centering
	\ra{1.1}
	\begin{tabular}[h]{@{} c c c c c c c  @{}}\toprule
		&$ $& $\Kp$ & $\Tn$ & $\Kv$  &$\fp$&$\fs$\\ 
		&$\mBO$ & $\!\times10^2$ & $\!\times10^2$ & $\!\times10^2$ & $\!\times10^3$  & $\!\times10^3$ \\ \midrule
		Nominal (no load) & - & 200 & 100 & 25 & 523 & 271\\
		Perf.-based (no load) & 67 & 445 & 125 & 40 & 401.9 & 224.9 \\
		Perf.-based (extra load) & 43 & 650 & 130 & 42.5 & 254.2 & 281.8 \\
		\bottomrule 
	\end{tabular}
\caption{Summary of all tuning results on the linear drive 
	}	
\label{table:expResults}
\end{table}

\begin{figure}[h]
	\centering
    \includegraphics[width=0.35\textwidth]{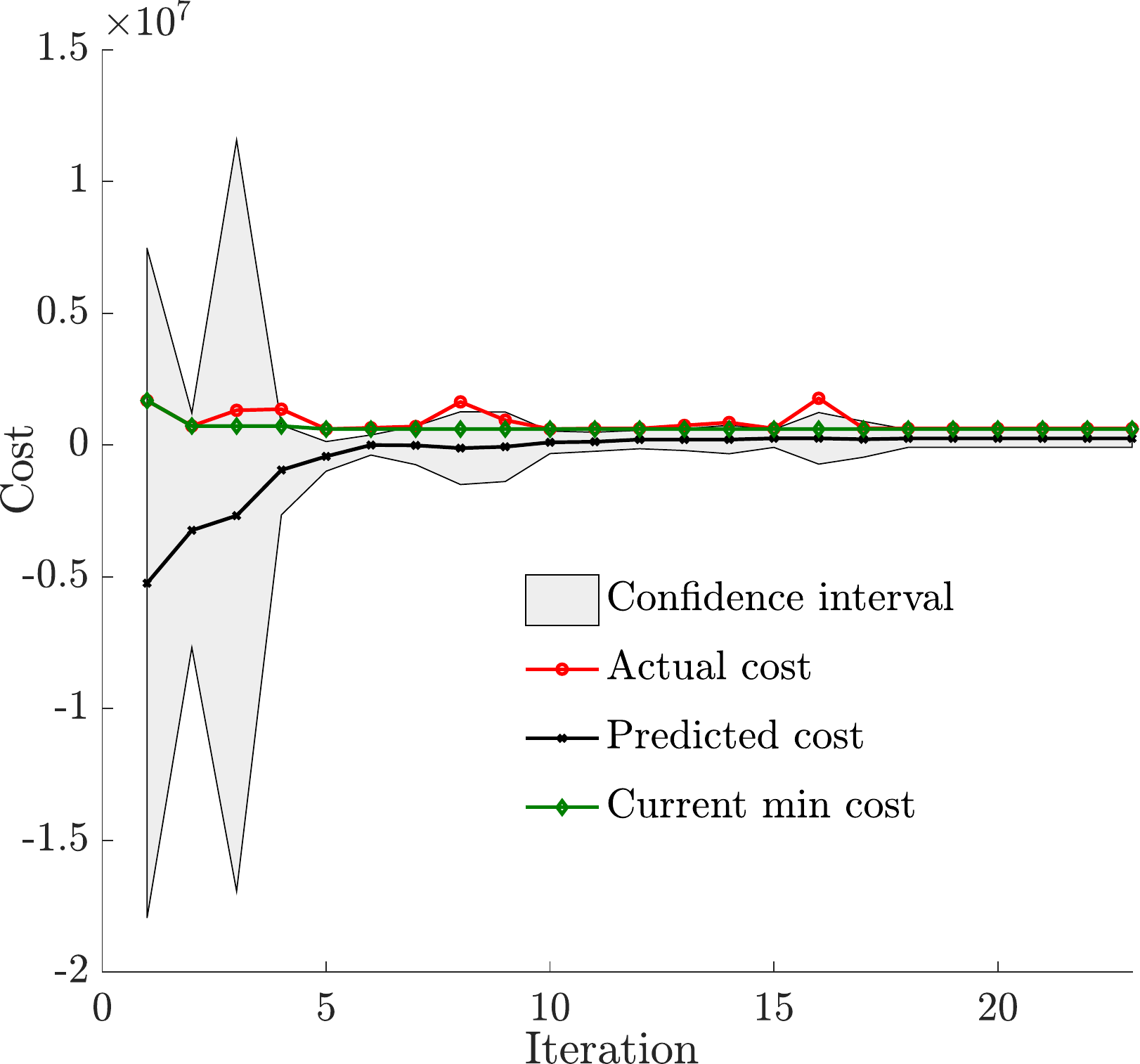}
      \caption{Predicted,   actual   observed   cost,   and   associated confidence interval at  3  standard  deviations  for  performance-based  BO  tuning on the system.}	\vspace{-5mm}
  \label{fig:BOexpCost_evolution}
\end{figure}

Because of the large number of samples in the training phase, the cost evolution during the BO iterations starts from a very low estimated cost, but with high uncertainty, and moves to higher cost, where the uncertainty is reduced after $10$ iterations, as shown in Figure \ref{fig:BOexpCost_evolution}. Looking at the {\em current minimal} cost, it can be seen that the BO-based tuning reaches a low-uncertainty cost which is very close to the converged results already at iteration 5. The algorithm converges after 18 iterations, according to the specified somewhat conservative termination criterion.

\begin{figure}[b]
\vspace{-4mm}
	\centering
    \includegraphics[width=0.475\textwidth]{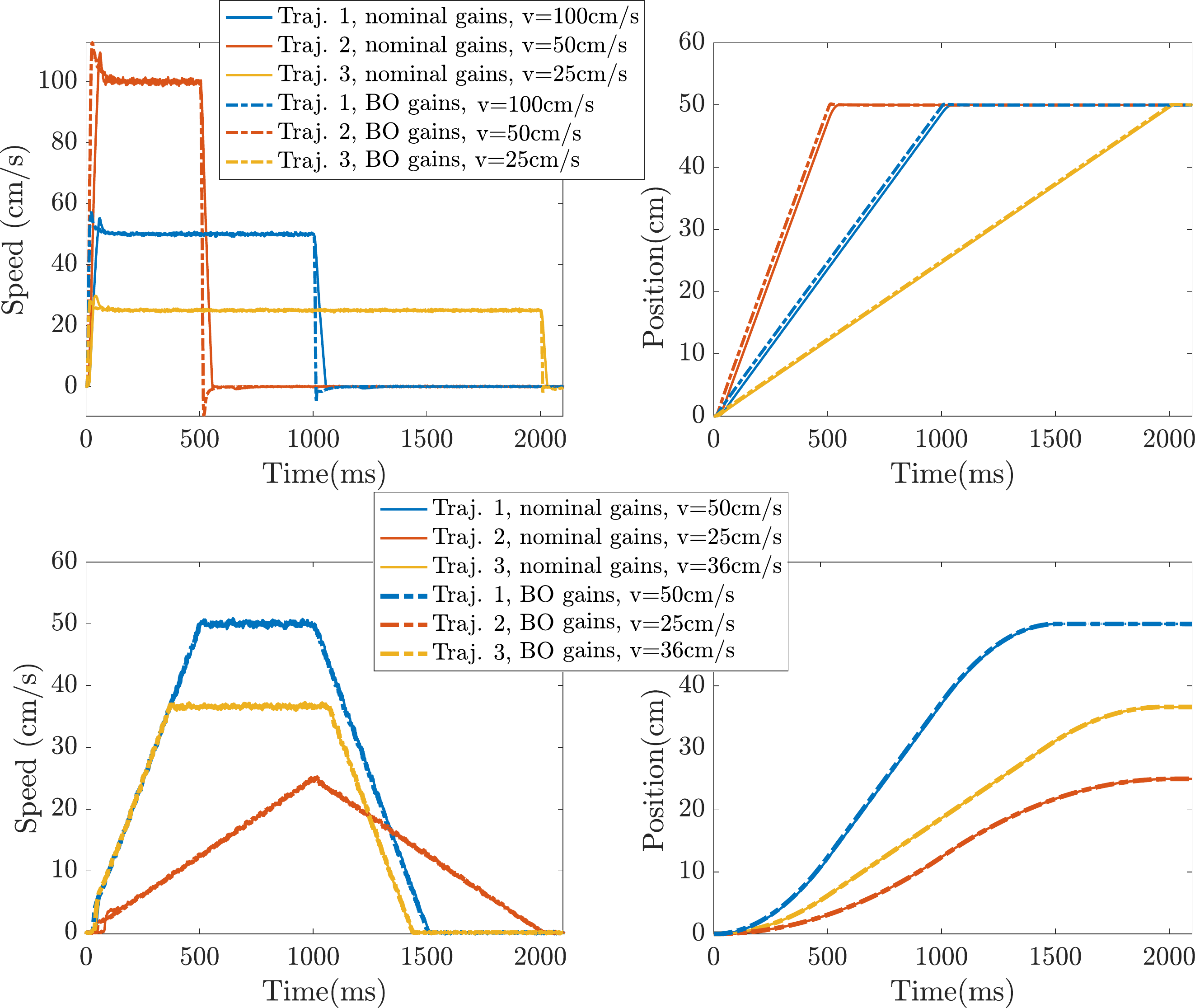}
      \caption{Position and speed responses comparing tracking with different trajectories. Upper panels: Position and speed responses for different step signals; Bottom panels: Trajectories with varying speed setpoints. 
      }
  \label{fig:BOExpTraj}
\end{figure}

We have further tested the robustness of the tuning approach by providing reference trajectories with different profiles. Figure \ref{fig:BOExpTraj} shows that the performance achieved with the BO tuning exceeds the nominal performance when there are fast changes in the speed, as in the top panels, and achieves better tracking of the references. The speed undershoot is significantly reduced, and better position stability is achieved at standstill, which is an important performance requirement for linear motion systems. For the trajectories shown in the bottom panels of Figure \ref{fig:BOExpTraj} the performance is virtually the same as nominal. 

\begin{figure}[b]
	\centering
	\vspace{-4mm}
	\includegraphics[width=0.45\textwidth]{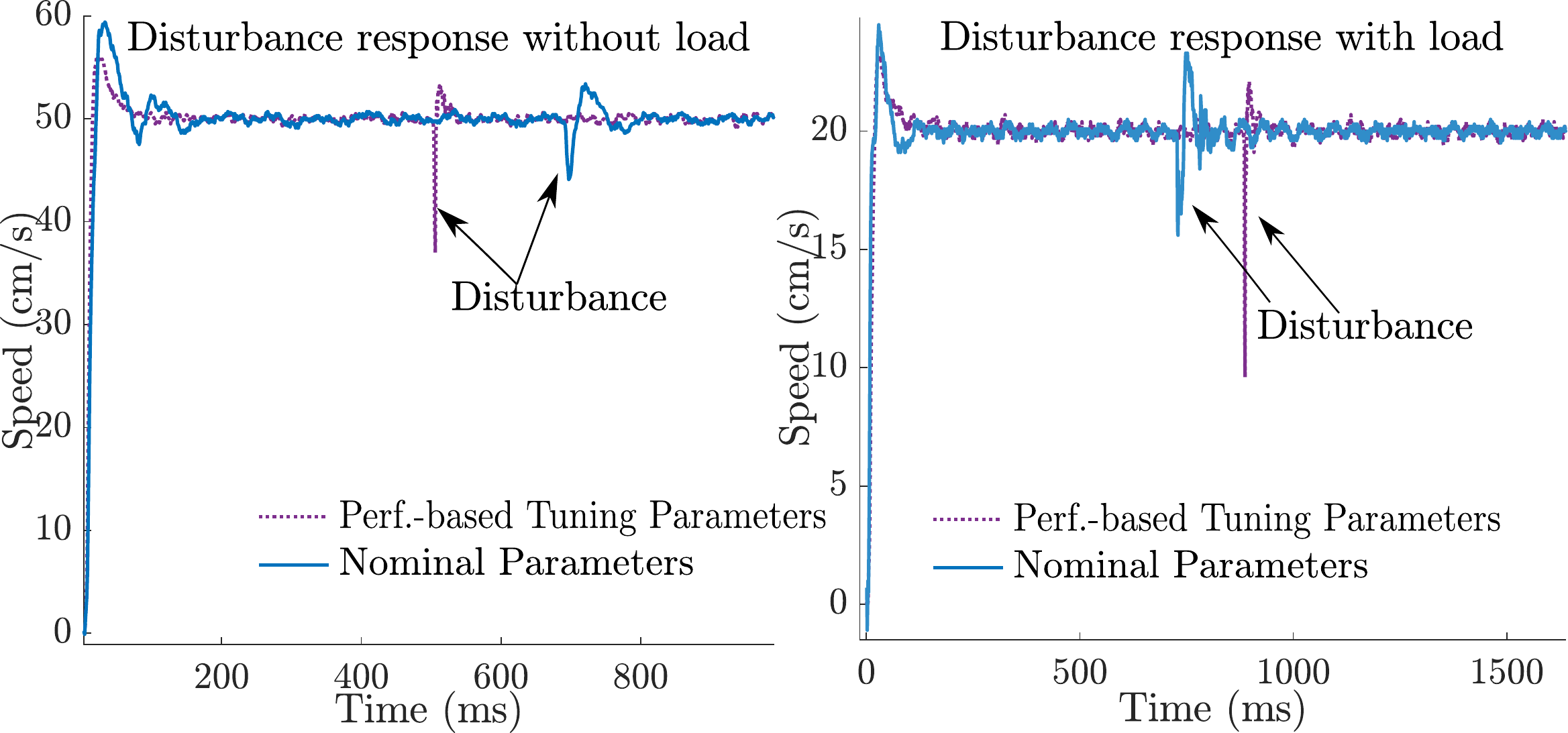}
	\caption{Speed responses of the system without and additional load, comparing disturbance response of the nominal controller and the BO-optimized controller.}
	\label{fig:expDisturb}
\end{figure}

Standard tuning methods such as the Ziegler-Nichols rule or relay tuning show excellent performance for disturbance rejection objectives. We have evaluated the BO-based controller parameters with respect to disturbance rejection. Impulsive disturbance forces were applied during the operation in the direction of movement of the system with additional load. The resulting response is shown in Figure \ref{fig:expDisturb}, and compared with the system's nominal performance when subjected to the same type of disturbance. The response of the system shows a quick recovery following the disturbance, as shown in Figure \ref{fig:expDisturb}. Even though the applied disturbance has significantly higher magnitude than in the nominal case, the observed recovery is faster for the performance-based tuned controller gains. This is due to the performance metric term that corresponds to tracking accuracy and position stability.

\subsection{Discussion} 
The proposed performance-based BO tuning thus offers a trade-off between grid based search, and heuristic-based methods. With BO tuning a relatively small number of experiments leads to the optimal gains, specified according to the data-driven optimization objective and termination criterion. The number of experiments can be varied, and usually prior experiments can be significantly reduced while achieving good performance.
Currently, tuning is performed in specified ranges to avoid destabilizing values of the parameters. The ranges are derived either based on the model of the system or based on expert knowledge from operating the system. When the ranges are not a priori known, and for the full automation of the method, it would be useful to ensure a safety mechanism in the performance-based tuning procedure. This could be achieved either through the addition of the safety constraints such as Gaussian processes in the BO algorithm or through the modification of the acquisition function to account for the probability of constraint violation \cite{Maier2019}. Another possibility is to detect experimentally when the system is approaching critical regime and to include a safety penalty in the cost \cite{SafetyBO2020}.
\section{Conclusion and Outlook} \label{sec:conclusion}
We have presented a data-driven approach for cascade controller tuning, where we model the performance of a closed-loop system as a function of controller gains. We apply a Bayesian optimization approach to derive controller gains with optimal performance. The performance was first evaluated in a simulation for a ball-screw linear axial system,  and compared to classical tuning approaches and the computed optimal performance on a grid. 
The experimental validation of the proposed method shows that it enables fast and standardized tuning, with a performance superior to other autotuning approaches. 
It allows easy adaptation of the controller parameters upon changes in the load or the system's mechanical configuration. Extending the method with automatic detection of instabilities will further increase its flexibility and potential for practical use.


\bibliography{bibliography} 

\begin{thebibliography}{10}
\providecommand{\url}[1]{#1}
\csname url@samestyle\endcsname
\providecommand{\newblock}{\relax}
\providecommand{\bibinfo}[2]{#2}
\providecommand{\BIBentrySTDinterwordspacing}{\spaceskip=0pt\relax}
\providecommand{\BIBentryALTinterwordstretchfactor}{4}
\providecommand{\BIBentryALTinterwordspacing}{\spaceskip=\fontdimen2\font plus
\BIBentryALTinterwordstretchfactor\fontdimen3\font minus
  \fontdimen4\font\relax}
\providecommand{\BIBforeignlanguage}[2]{{%
\expandafter\ifx\csname l@#1\endcsname\relax
\typeout{** WARNING: IEEEtran.bst: No hyphenation pattern has been}%
\typeout{** loaded for the language `#1'. Using the pattern for}%
\typeout{** the default language instead.}%
\else
\language=\csname l@#1\endcsname
\fi
#2}}
\providecommand{\BIBdecl}{\relax}
\BIBdecl

\bibitem{feedbackSys}
K.~J. Astr{\"o}m and R.~M. Murray, \emph{Feedback Systems: An Introduction for
  Scientists and Engineers}.\hskip 1em plus 0.5em minus 0.4em\relax USA:
  Princeton University Press, 2008.

\bibitem{kessler58}
C.~Kessler, ``{Das symmetrische Optimum},'' \emph{Regelungstechnik}, vol.~6,
  pp. 395 -- 400, 432 -- 436, 1958.

\bibitem{Preitl99}
S.~Preitl and R.-E. Precup, ``An extension of tuning relations after
  symmetrical optimum method for {PI} and {PID} controllers,''
  \emph{Automatica}, vol.~35, no.~10, pp. 1731 -- 1736, 1999.

\bibitem{Karimi2017}
A.~Karimi and C.~Kammer, ``A data-driven approach to robust control of
  multivariable systems by convex optimization,'' \emph{Automatica}, vol.~85,
  pp. 227 -- 233, 2017.

\bibitem{daSilva2000}
W.~G. {da Silva}, P.~P. {Acarnley}, and J.~W. {Finch}, ``Application of genetic
  algorithms to the online tuning of electric drive speed controllers,''
  \emph{IEEE Transactions on Industrial Electronics}, vol.~47, no.~1, pp.
  217--219, 2000.

\bibitem{Qi2019}
Z.~{Qi}, Q.~{Shi}, and H.~{Zhang}, ``Tuning of digital {PID} controllers using
  particle swarm optimization algorithm for a {CAN}-based {DC} motor subject to
  stochastic delays,'' \emph{IEEE Transactions on Industrial Electronics},
  vol.~67, no.~7, pp. 5637--5646, 2020.

\bibitem{ziegler1942optimum}
J.~Ziegler and N.~Nichols, ``Optimum settings for automatic controllers,''
  \emph{Transactions of the ASME}, vol.~64, pp. 759--768, 1942.

\bibitem{hang2002relay}
C.~Hang, K.~{\AA}str{\"o}m, and Q.~Wang, ``Relay feedback auto-tuning of
  process controllers--a tutorial review,'' \emph{Journal of process control},
  vol.~12, no.~1, pp. 143--162, 2002.

\bibitem{Kumar2015}
R.~Kumar, S.~Singla, and V.~Chopra, ``Comparison among some well known control
  schemes with different tuning methods,'' \emph{Journal of Applied Research
  and Technology}, vol.~13, no.~3, pp. 409 -- 415, 2015.

\bibitem{Campi2002}
M.~Campi, A.~Lecchini, and S.~Savaresi, ``Virtual reference feedback tuning: a
  direct method for the design of feedback controllers,'' \emph{Automatica},
  vol.~38, no.~8, pp. 1337 -- 1346, 2002.

\bibitem{Campi2006}
M.~C. {Campi} and S.~M. {Savaresi}, ``Direct nonlinear control design: the
  virtual reference feedback tuning ({VRFT}) approach,'' \emph{IEEE
  Transactions on Automatic Control}, vol.~51, no.~1, pp. 14--27, 2006.

\bibitem{Novara2018}
C.~{Novara} and S.~{Formentin}, ``Data-driven inversion-based control of
  nonlinear systems with guaranteed closed-loop stability,'' \emph{IEEE
  Transactions on Automatic Control}, vol.~63, no.~4, pp. 1147--1154, 2018.

\bibitem{Formentin2016}
S.~Formentin, D.~Piga, R.~Toth, and S.~M. Savaresi, ``Direct learning of {LPV}
  controllers from data,'' \emph{Automatica}, vol.~65, pp. 98 -- 110, 2016.

\bibitem{Prochazka2005}
H.~{Prochazka}, M.~{Gevers}, B.~D.~O. {Anderson}, and C.~{Ferrera}, ``Iterative
  feedback tuning for robust controller design and optimization,'' in
  \emph{Conference on Decision and Control}, pp. 3602--3607, 2005.

\bibitem{Radac2014}
M.~{Rădac}, R.~{Precup}, E.~M. {Petriu}, and S.~{Preitl}, ``Iterative
  data-driven tuning of controllers for nonlinear systems with constraints,''
  \emph{IEEE Transactions on Industrial Electronics}, vol.~61,
  \href{http://dx.doi.org/10.1109/TIE.2014.2300068}{DOI
  10.1109/TIE.2014.2300068}, no.~11, pp. 6360--6368, Nov. 2014.

\bibitem{Bazanella2014}
A.~S. Bazanella, L.~Campestrini, and D.~Eckhard, \emph{Data-Driven Controller
  Design: The $H_2$ Approach}.\hskip 1em plus 0.5em minus 0.4em\relax Springer
  Publishing Company, Incorporated, 2014.

\bibitem{Tomlin}
S.~{Bansal}, R.~{Calandra}, T.~{Xiao}, S.~{Levine}, and C.~J. {Tomiin},
  ``Goal-driven dynamics learning via {B}ayesian optimization,'' in
  \emph{Conference on Decision and Control}, pp. 5168--5173, 2017.

\bibitem{Maier2019}
M.~Maier, R.~Zwicker, M.~Akbari, A.~Rupenyan, and K.~Wegener, ``Bayesian
  optimization for autonomous process set-up in turning,'' \emph{CIRP Journal
  of Manufacturing Science and Technology}, vol.~26,
  \href{http://dx.doi.org/https://doi.org/10.1016/j.cirpj.2019.04.005}{DOI
  https://doi.org/10.1016/j.cirpj.2019.04.005}, pp. 81 -- 87, 2019.

\bibitem{Maier_2020gr}
M.~Maier, A.~Rupenyan, C.~Bobst, and K.~Wegener, ``Self-optimizing grinding
  machines using gaussian process models and constrained bayesian
  optimization,'' \emph{The International Journal of Advanced Manufacturing
  Technology}, vol. 108, pp. 528--552, 2020, {D}OI:
  {https://doi.org/10.1007/s00170-020-05369-9}.

\bibitem{Andersson2016}
O.~{Andersson}, M.~{Wzorek}, P.~{Rudol}, and P.~{Doherty}, ``Model-predictive
  control with stochastic collision avoidance using {B}ayesian policy
  optimization,'' in \emph{2016 IEEE International Conference on Robotics and
  Automation (ICRA)}, \href{http://dx.doi.org/10.1109/ICRA.2016.7487661}{DOI
  10.1109/ICRA.2016.7487661}, pp. 4597--4604, May. 2016.

\bibitem{BayesOpt1}
F.~Berkenkamp, A.~P. Schoellig, and A.~Krause, ``Safe controller optimization
  for quadrotors with {G}aussian processes,'' \emph{2016 IEEE International
  Conference on Robotics and Automation}, pp. 491 -- 496, 2016.

\bibitem{BayesOpt2}
M.~Neumann-Brosig, A.~Marco, D.~Schwarzmann, and S.~Trimpe, ``Data-efficient
  auto-tuning with {B}ayesian optimization: {A}n industrial control study,''
  \emph{IEEE Transactions on Control Systems Technology}, 2018.

\bibitem{khosravi2019controller}
M.~Khosravi, A.~Eichler, N.~Schmid, P.~Heer, and R.~S. Smith, ``Controller
  tuning by {B}ayesian optimization an application to a heat pump,'' in
  \emph{European Control Conference}, pp. 1467--1472.\hskip 1em plus 0.5em
  minus 0.4em\relax IEEE, 2019.

\bibitem{khosravi2019machine}
M.~Khosravi, N.~Schmid, A.~Eichler, P.~Heer, and R.~S. Smith, ``Machine
  learning-based modeling and controller tuning of a heat pump,'' in
  \emph{Journal of Physics: Conference Series}, vol. 1343, no.~1, p.
  012065.\hskip 1em plus 0.5em minus 0.4em\relax IOP Publishing, 2019.

\bibitem{BayesOpt5}
F.~Berkenkamp, A.~Krause, and A.~P. Schoellig, ``{B}ayesian optimization with
  safety constraints: Safe and automatic parameter tuning in robotics,''
  \emph{arXiv:1602.0445}, 2016.

\bibitem{SafetyBO2020}
C.~Koenig, M.~Khosravi, R.~S. Smith, A.~Rupenyan, and J.~Lygeros,
  ``Safety-aware cascade controller tuning using constrained {B}ayesian
  optimization,'' \emph{Master thesis, ETH Zurich}, 2020.

\bibitem{altintas2011machine}
Y.~Altintas, A.~Verl, C.~Brecher, L.~Uriarte, and G.~Pritschow, ``Machine tool
  feed drives,'' \emph{CIRP annals}, vol.~60, no.~2, pp. 779--796, 2011.

\bibitem{GPR}
C.~E. Rasmussen and C.~K.~I. Williams, \emph{Gaussian Processes for Machine
  Learning}.\hskip 1em plus 0.5em minus 0.4em\relax MIT Press, 2006.

\bibitem{srinivas2012information}
N.~Srinivas, A.~Krause, S.~M. Kakade, and M.~W. Seeger, ``Information-theoretic
  regret bounds for {G}aussian process optimization in the bandit setting,''
  \emph{IEEE Transactions on Information Theory}, vol.~58, no.~5, pp.
  3250--3265, 2012.

\bibitem{AxisModelling}
K.~K. Varanasi and S.~A. Nayfeh, ``The dynamics of lead-screw drives: Low-order
  modeling and experiments,'' \emph{Journal of Dynamic Systems, Measurement,
  and Control, ASME}, vol. 126, pp. 388 -- 398, Jun. 2004.

\bibitem{2DOFControl1}
R.~Qian, M.~Luo, J.~Zhao, and T.~Li, ``Novel sliding mode control for ball
  screw servo system,'' in \emph{MATEC Web of conferences,7th International
  Conference on Mechanical, Industrial, and Manufacturing Technologies}, ser.
  03007, vol.~54, 2016.

\bibitem{aastrom1993automatic}
K.~J. {\AA}str{\"o}m, T.~H{\"a}gglund, C.~C. Hang, and W.~K. Ho, ``Automatic
  tuning and adaptation for {PID} controllers--a survey,'' \emph{Control
  Engineering Practice}, vol.~1, no.~4, pp. 699--714, 1993.

\bibitem{IFACBOCascade}
M.~Khosravi, V.~Behrunani, R.~S. Smith, A.~Rupenyan, and J.~Lygeros, ``Cascade
  control: Data-driven tuning approach based on {B}ayesian optimization,''
  \emph{2020 IFAC World Congress (arXiv:2005.03970)}, 2020.

\end{thebibliography}
\bibliographystyle{IEEEtranTIE}
\end{document}